\documentclass[sigconf,nonacm]{acmart}

\usepackage{amsfonts,amssymb,amsmath}
\usepackage{algorithmic}
\usepackage{graphicx}
\usepackage{textcomp}
\usepackage{xcolor}
\usepackage{soul}

\usepackage{svg}
\newcommand{\orcidi}[1]{\href{https://orcid.org/#1}{\includegraphics[width=8pt]{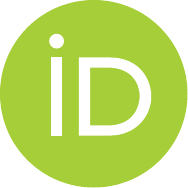}}}

\let\svthefootnote\thefootnote
\newcommand\freefootnote[1]{%
  \let\thefootnote\relax%
  \footnotetext{#1}%
  \let\thefootnote\svthefootnote%
}

\usepackage{balance}
\usepackage{verbatim}
\usepackage{enumitem}
\usepackage{hyperref}

\usepackage{xcolor}
\usepackage{soul}

\hypersetup{hidelinks}

\usepackage{algorithm}

\begin{document}

\title{Extending the RISC-V ISA for exploring \\ advanced reconfigurable SIMD instructions}
\author{Philippos Papaphilippou \orcidi{0000-0002-7452-7150}, Paul H. J. Kelly, Wayne Luk}

\affiliation{%
  \country{Department of Computing, Imperial College London, UK}
}
\email{{pp616, p.kelly, w.luk}@imperial.ac.uk}%

\begin{abstract}
This paper presents a novel, non-standard set of vector instruction types for exploring custom SIMD instructions in a softcore. The new types allow simultaneous access to a relatively high number of operands%
, reducing the instruction count where applicable.
Additionally, a high-performance open-source
RISC-V (RV32 IM) softcore is introduced, optimised for exploring custom SIMD instructions and streaming performance. 
By providing instruction templates for instruction development in HDL/Verilog, efficient FPGA-based instructions can be developed with few low-level lines of code. 
In order to improve custom SIMD instruction performance, the
softcore's cache hierarchy is optimised for bandwidth,
such as with very wide blocks for the last-level cache. The approach
is demonstrated on example memory-intensive applications on an FPGA. %
Although the exploration is based on the softcore, the goal is to provide a means to experiment with advanced SIMD instructions which could be loaded in future CPUs that feature reconfigurable regions as custom instructions.
Finally, we provide some insights on the challenges and effectiveness of such future micro-architectures. %

\freefootnote{CARRV 2021, June 17, Virtual Workshop, Co-located with ISCA 2021\vspace{-3.42em}}

\end{abstract}
\maketitle
\fancyhead[LE,RO]{\textsf{\footnotesize CARRV 2021, June 17, 2021, Virtual Workshop}}
\fancyhead[RE]{\textsf{\footnotesize Philippos Papaphilippou, Paul H. J. Kelly, Wayne Luk}}

\section{Introduction}

Modern general purpose processors (CPUs) support a wide range of single-instruction-multiple data (SIMD) instructions \cite{intel} as a way to accelerate applications that exhibit data-level parallelism. While there can be notable performance gains in certain applications \cite{bramas}, the instruction set extensions become bloated with overspecialised instructions \cite{flexbex}, and sometimes it is difficult to express efficiently a parallel task using a fixed set of SIMD extensions \cite{chhugani2008efficient}.

As an alternative means to acceleration, FPGAs %
achieve unparalleled processing capabilities in specialised tasks \cite{nakahara2020high,tatsumura2019fpga}. They have been getting attention for datacenter use, with numerous academic and industrial solutions focusing on processing big data \cite{owaida2017centaur, kara2019doppiodb, psurvey}. However, the combination of specialisation and their placement in highly heterogeneous systems has high development and deployment costs.

FPGAs are often left behind in terms of main memory bandwidth, leading to a bandwidth bottleneck for big data accelerators \cite{casper2014hardware,owaida2017centaur,halstead2015fpga}. Also, high-end FPGAs are usually based on non-uniform memory access (NUMA) systems, and the communication techniques are mostly inconvenient and expensive: First, PCIe, the most widely-used interconnection protocol has a high access latency \cite{stuecheli2018ibm}, and is not appropriate for low-latency applications. Second, vendors promote high-level synthesis (HLS) tools in an effort to abstract the complexity of communication, at the expense of enforcing generalised programming models. %
Last, unlike CPUs, %
any cache memory hierarchy, such as for improving random memory accesses, is usually implemented from scratch on FPGAs, and this complexity is often reflected in designs \cite{halstead2015fpga}.

In combination with the openness of the RISC-V instruction set, and its support for custom instructions  \cite{waterman2020risc}, this is a great time to start considering custom SIMD instructions on general purpose CPUs. Small FPGAs can be integrated, to implement custom instructions \cite{flexbex} and are demonstrated to improve the performance over existing extensions significantly \cite{ordaz2016soft,ordaz2018soft}. In the literature, the exploration of custom instructions on CPUs, and specifically SIMD, is rather limited, even though small reconfigurable regions working as instructions is a promising idea and possibly the future of acceleration.

In this paper, we present novel instruction types, Verilog templates and an open-source framework that are optimised for exploring custom SIMD instructions. In order to achieve high throughput on the provided softcore for streaming applications, the focus was given on the cache hierarchy and communication. The framework allows easy integration of custom vector instructions and evaluation in simulation and hardware. %
Finally, we evaluate examples of custom instructions and provide insights on introducing small FPGAs as execution units in CPUs. %
Our contributions are as follows:

\begin{itemize}[leftmargin=2em]

\item A set of experimental non-standard instruction types to enable optional access to a high number of registers for expressing complex tasks with fewer instructions (section \ref{vit}).
\item An open-source\footnote{\emph{Source available:} \url{http://philippos.info/simdsoftcore} \vspace{-3.42em}%
} softcore framework to evaluate novel SIMD custom instructions, and design choices to maximise streaming performance (sections \ref{softc},\ref{dse}).
\item Defining a clean Verilog template for custom acceleration as an SIMD ISA extension in a lightweight softcore (section \ref{instt}).%
\item A demonstration of the approach with novel SIMD instructions for sorting and prefix sum (section \ref{simdu}). %
\end{itemize}

\section {Custom SIMD instructions}\label{simdr}

In order to support adding and using SIMD instructions on the proposed softcore, we introduce \emph{(A)} new instruction types that refer to the vector registers and \emph{(B)} HDL (Verilog) code templates for implementing the instructions in hardware. %

\subsection{Vector instruction types}\label{vit}

In addition to the RV32IM standards we propose two additional non-standard instruction types for supporting the custom SIMD instructions. Originally there are 4 main instruction types: R, I, S/B, U/J in RV32I, that define the  format of the immediate field and the operands. The proposed instruction types I' and S' are variations of the I and S types respectively. 

The official draft RISC-V "V" vector extension %
has not been followed in this work, as it seems to target high-end/ hardened processors. %
For example, it requires a fixed number of 32 vector registers and features hundreds of instructions \cite{rvv}, also reducing the number of registers an instruction can access. %
For our use case, this would be contrary to the idea of having small reconfigurable regions as instructions, rather than supporting hundreds of intrinsics.

The modification repurposes the space used for the immediate field for the vector register names. \textbf{Up to 6 registers} (vector and non-vector) can be accessed by the same instruction, reducing the instruction count for complex or stateful (through registers) instructions. The use of the immediate field for the vector registers was also convenient for minimal interference and modification to the RISC-V GNU toolchain in GCC's binutils for supporting inline assembly of custom instructions. There is currently no official tool support for vector registers due to the draft status of the vector extension. We opted to use the opcodes dedicated to custom applications for all custom vector instructions, for inline assembly.

\begin{figure}[h!]
\centering
\includegraphics[width=0.5\textwidth, trim=0 10 -20 0]{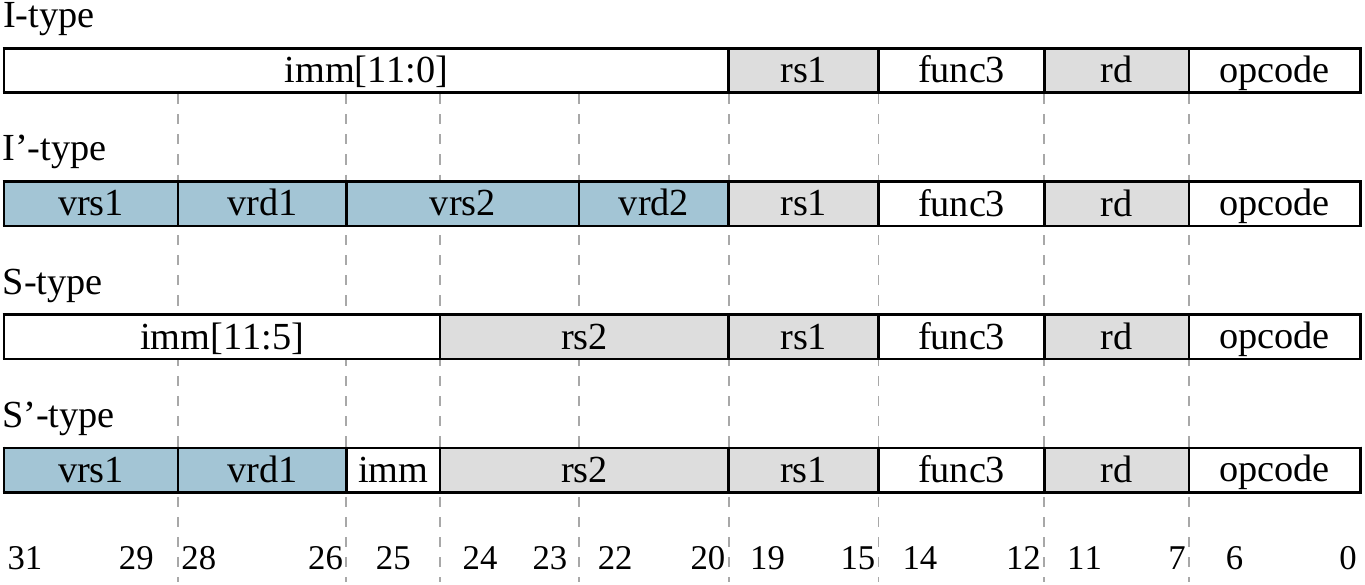}
\caption{Two variations of the I and S instruction types}\label{fity}
\vspace{-0.5em}
\end{figure}

Type I' provides access to the register operands of the I-type, that is one source and one destination 32-bit register. It also provides access to two source (\emph{vrs1} and \emph{vrs2}) and two destination (\emph{vrd1} and \emph{vrd2}) vector registers. Type S' exchanges the space used by \emph{vrs2} and \emph{vrd2} for access to an additional 32-bit source register \emph{rs2}. The latter would be useful, for example, for load and store instructions, for breaking loop indexes into two registers and potentially reducing the instruction count in some loops, as in other ISAs. The proposed variations are summarised in Figure \ref{fity}.

As shown in Figure \ref{fity}, the fields for each vector register are three bits wide, which sets the maximum number of vector registers to 8. Vector register 0 is conveniently assigned to a \textbf{constant value of 0}, similarly to the 32-bit base registers. It is useful for the proposed many-register instructions types, because not all register operands may need to be accessed at once, allowing different combinations of operands using the same type. In software, this can be achieved with aliasing the unused operands with register 0, and was \textbf{not} supported in ``V'', as vector 0 originally represents a register.

An interesting feature in the ``V'' specification is the ability to chain vector instructions, hence the need for high number of registers. Since our solution enables custom instruction \textbf{pipelines of arbitrary length}, a lower number of registers was considered satisfactory, as the \textbf{need for chaining is minimised}. %

\subsection{Instruction templates}\label{instt}

The custom instruction templates are placeholder modules inside the softcore codebase, for adding user code for specialised SIMD instruction implementations. Algorithm \ref{templa} is written in Verilog and shows part of the template for  I'-type instructions, plus an example user code, marked in yellow. On each cycle, the instruction module also accepts the destination register names (\emph{rd, vrd1} and \emph{vrd2}) to provide them later, when the data result is ready, after the specified pipeline length (\emph{c1\_cycles}). In this way, with a pipelined implementation the module can process \textbf{multiple calls} one after another. %
Blocking instructions are also supported with minor modification.
\begin{algorithm}[h]
\centering
\includegraphics[width=0.50\textwidth, trim=0 -5 -5 0]{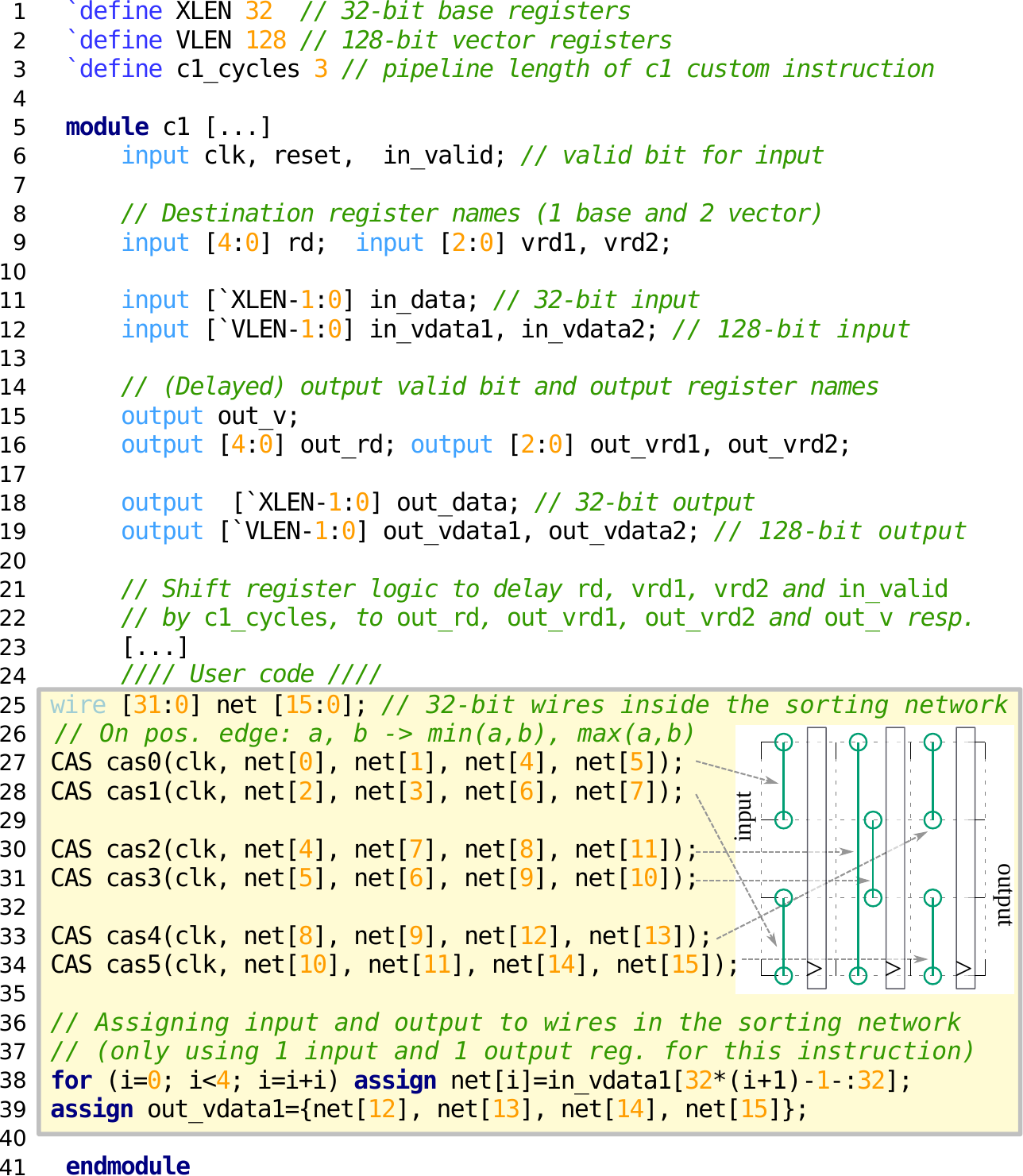}
\caption{Verilog template for I'-type, with user-provided code highlighted in yellow}\label{templa}%

\end{algorithm}

The example instruction implementation in Algorithm \ref{templa} is a bitonic sorter of 4 inputs. Such sorting networks are parallel and pipelinable algorithms for sorting a list of \(N\) values. In each parallel step there is a number of compare-and-swap (CAS) units, that collectively sort the entire input list, as the input moves along the network. The odd-even merge sorter and the bitonic sorter \cite{batcher1968sorting} are two similar sorting network topologies, both consisting of \(\Theta(log^2(N))\) parallel steps. For a vector register width of 128-bit, this bitonic sorter sorts four 32-bit values in 3 cycles.

The template of the S'-type instructions is similar to Algorithm \ref{templa}, but with an interface that reflects the correct operands (2 base and 1 vector registers for input and 1 base and 1 vector for output). One S' type instruction for loading and storing VLEN-sized vectors (\emph{c0\_lv} and \emph{c0\_sv} respectively) is provided by default.

\section{Softcore microarchitecture}\label{softc}

The proposed softcore supports the RISCV 32-bit base integer instruction set (RV32I v. 2.1), plus the ``M'' extension for integer multiplication and division \cite{waterman2020risc}. The novel features of our approach include a series of design choices to: (1) enable high-performance for custom vector instructions and streaming applications; (2) allow efficient implementation on recent FPGAs by enhancing the block RAM (BRAM) organisation and the behaviour of the inter-chip communication. %

\subsection {Cache hierarchy optimisations}

\begin{figure}[h!]
\centering
\includegraphics[width=0.35\textwidth, trim=0 0 0 20]{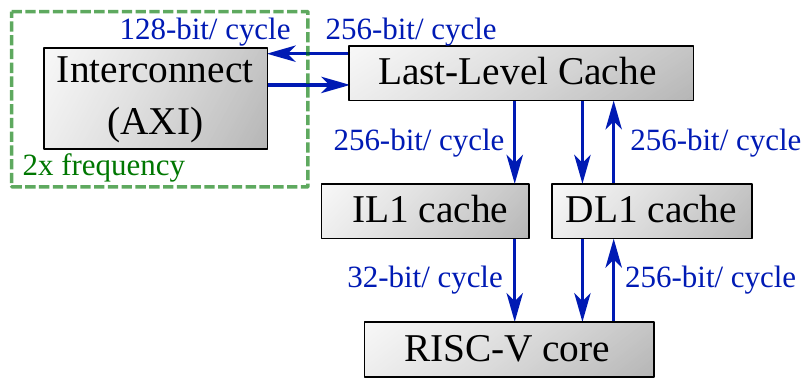}
\caption{Data movement in an example cache configuration}\label{hierf}
\vspace{-1em}
\end{figure}

On the first level, there is an instruction cache (IL1) and a data cache (DL1), and on the second level there is a unified last-level cache (LLC). LLC responds to requests from both IL1 and DL1.%
It communicates in bursts to DRAM thorough an interconnect such as AXI. It resembles a modified Harvard architecture, as the address space is common between data and instructions. %
Figure \ref{hierf} provides a high-level example for the data communication throughout the cache hierarchy.\\

All caches use the writeback policy, with the exception of the IL1 cache, where writing is not needed. This is achieved by storing a dirty bit alongside each stored block, to acknowledge modification. While DL1  and LLC are  set-associative caches, the IL1 is direct-mapped for fast lookup of the next instruction, to avoid a stall on instruction hits. 

The LLC is implemented in block memory (BRAM), to accommodate the high capacity of the last level. The IL1 is implemented in registers for a reduced latency, in order to provide the successor instruction immediately on the next cycle (and avoiding being the critical path) on hits. The DL1 is implemented in BRAM by default, although changing the directive to registers yields similar performance and utilisation results, due to its relatively low size. 

At the set-associative caches (DL1 and LLC), each block can be allocated into multiple possible ways, represented by different parallel block RAM sections. The block replacement policy for these caches
is not-recently-used (NRU). It uses one bit of meta-information per block \cite{ultrasparc2006supplement}, but closely resembles the Least-Recently-Used (LRU) eviction policy. %
The choice of a  replacement policy can be crucial to the performance of streaming applications, due to the wide data blocks and the reduced cache space on the FPGA. For instance, a random policy would stagnate the bandwidth for memory copying (\emph{memcpy()}), when the source and destination are aligned. %

A series of design choices are presented for optimising the performance and applicability for our purposes. %

\subsubsection{Level-1 block size}

One optimisation is to set the block size of the DL1 to be equal to the \textbf{vector register width}, such as 256 bits. This is because a wider block size would require an additional read on each write, from the cache of the higher level, so that the entire block becomes valid. When the data are from vector registers and are properly aligned, there is no need to wait for fetching that block on a write miss, because the whole block will contain new information. %

The IL1 uses the same block size for easier arbitration between DL1 blocks, at the cache of the higher-level (LLC). Additionally, since IL1 is direct-mapped, using a wider-block than 32-bits is also beneficial to performance, as it can also be seen as a natural way of prefetching.

\subsubsection{LLC block size}

An important feature for increased streaming performance is \textbf{very wide blocks} for LLC, such as 8192-bit wide. This is in contrary to today's CPUs with a 512-bit (64-byte) block size. The idea is that on write-back to/ fetch from main memory, it achieves a higher speed because of longer bursts. Longer bursts are shown to have significant impact on the overall throughput, such as in heterogeneous systems with AXI \cite{manev2019unexpected}, and this is especially useful for streaming. Associating entire LLC blocks with bursts was a convenient and practical organisation choice, because of interconnect protocol limitations, such as for not crossing the 4KB address boundary in AXI \cite{holdings2013amba}.

\subsubsection{LLC strobe functionality}

A naive implementation of the LLC in BRAM, would be to read the (wide) blocks in their entirety in a single cycle, as in the DL1 case. However, BRAM is organised in chunks of certain width and length, such as 36-bit wide. With a LLC of just a single wide-enough block, the BRAM capacity of the FPGA can be exceeded, or stagnate timing performance. 
For this reason, the proposed LLC stores blocks in  consecutive BRAM \textbf{locations of narrower size}.
There is an internal notion of sets that corresponds to the address of the block memory, where each requested data can reside. The tag array only stores the tags of entire blocks. %

There is no overhead in access latency by using sub-blocks, as it still takes a single cycle to read an I/DL1-sized block from LLC. %
Another advantage of this technique is that, on fetch, the requested I/DL1 block can be provided before a read burst from DRAM finishes, since the LLC blocks are stored progressively. %
\subsubsection{Doubling the frequency of the interconnect}

In contrast to the timing characteristics of this softcore, as well with other well-known softcores \cite{heinz2019catalog}, the operating frequency of the interconnect on FPGAs can be relatively much higher \cite{interconnect2017v2}. Given that the port data widths are rather narrow (e.g. 128 bits/ cycle), this directly impacts the throughput for streaming applications. This optional %
optimisation involves setting double rate for the interconnect, %
to \textbf{emulate double data width} by fetching or writing twice per cycle, and saturate \cite{manev2019unexpected} the bandwidth more easily (see Figure \ref{hierf}).

\subsection {Main core}

The core has a single pipeline stage, even though more advanced instructions such as pipelined vector instructions have their own pipeline. Almost all instructions in RV32I consume 1 cycle and the result is available on the immediately next cycle. In practice, this has a similar effect to operand forwarding in pipelined processors, as consecutive dependent instructions are executed sequentially without stalls. %

The load and store instructions are handled by the cache system independently. On a data cache hit there is a latency of 3 cycles until the dependent command gets executed. The 3 cycles can be seen as a small pipeline with one cycle for memory access, one for fetching the data and one for updating the registers. This effectively yields a latency of 2 cycles for cache hits, when the next instruction is data-dependent on the load, as the execution is in-order.%

Having a \textbf{single pipeline stage}, so that most instructions complete in a single cycle, 
is useful for simplifying the dependency checks. The output of simple instructions such as add, addi, etc. is not tracked for dependencies. 
Of course, there are alternative approaches%
, but the current implementation mapped well in our evaluation platform and facilitated the SIMD functionality rather efficiently.

The implementation of the SIMD instructions follow the templates of section \ref{simdr}), that allow a variable pipeline length, abstracted through a \(ready\) signal for when the result is available. %
Apart from the 32 base 32-bit registers (as per RV32I), there are up to 8 VLEN-wide registers, such as 256-bit-wide for the SIMD instructions. In both sets of registers, the register 0 is driven by the constant 0. %

\section{Evaluation}\label{eval}

The exploration is divided in three parts according to the outcomes of each set of experiments: (\ref{dse}) justifies important design choices related to streaming performance, (\ref{rvev}) shows that the performance is still acceptable when no SIMD instruction is used and (\ref{simdu}) explores the behaviour and efficiency of example novel custom SIMD instructions.

The evaluation platform is Ultra96, which features the Xilinx UltraScale+ ZU3EG device. The FPGA on the device shares the same 2GB DDR4 main memory with the 4 ARM cores. The ARM cores run Linux, but the kernel address space is manually configured to end at the 1GB mark, so that the other 1GB is dedicated to the FPGA, 
that includes the softcore. %

\subsection{Design Space Exploration}\label{dse}

The target application is memory copying (\(memcpy()\)), %
as its performance is (indirectly) detrimental to big data processing and related evaluations have a long history in HPC applications \cite{mccalpin1995stream}. \(memcpy()\) here is manually implemented with the custom instructions for load vector and store vector, instead of a library implementation using base registers. The data length is 256 MiB, in order to %
surpass the cache sizes.

\begin{figure}[h!]
\centering
\includegraphics[width=0.5\textwidth, trim=0 0 -15 -5]{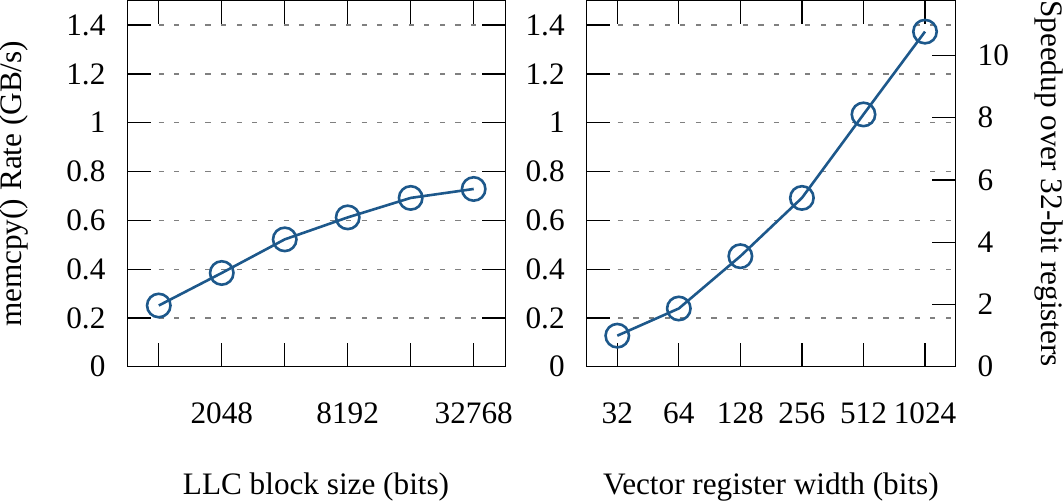}
\caption{Memcpy() read and write (bidirectional) throughput for different last-level cache block sizes (left) and vector register widths (right)}\label{memf}
\vspace{-0.5em}
\end{figure}

Figure \ref{memf} (right) illustrates the impact of the vector register size on \(memcpy()\). The 1024-bit softcore achieved a memcpy() rate of 1.37 GB/s. Though, we opt for 256-bit (VLEN) registers, with a rate of 0.69 GB/s, as 512-bit and beyond seemed more challenging to route efficiently when incorporating more complex custom instructions. %
These designs used a 16384-bit-wide LLC block.

One other experiment (Figure \ref{memf} left) measures the impact of the block size in last-level cache (LLC). %
Wider LLC blocks seem to be a considerable contributor to memory performance, as they relate to the burst size.  The improvement starts to plateau after longer bursts at around 8192 bits. All implementations reached timing closure for a frequency 150 MHz, except the 1024-bit configuration that %
was clocked at 125 MHz. 
Table \ref{tab1} summarises the selected baseline configuration for the remainder of the evaluation.

\vspace{-0.4em}
\begin{table}[h!] 
\footnotesize	
\caption{Selected configuration} 
\label{tab1}

\centering
\setlength{\tabcolsep}{3pt}
\begin{tabular} {c c | c c c | c c c c | c | c }
\multicolumn{2}{c|}{IL1}&\multicolumn{3}{c|}{DL1}&\multicolumn{4}{c|}{LLC}&&\\
sets&block&sets&ways&block&sets&ways&block&sub-blocks&VLEN&\(f_{max}\)\\
&(bits)&&&(bits)&&&(bits)&&(bits)&(MHz)\\
\hline
&&&&&&&&&&\\
64&256&32&4&256&32&4&16384&32&256&150\\
\multicolumn{2}{c|}{(=2KiB)}&\multicolumn{3}{c|}{(=4KiB)}&\multicolumn{4}{c|}{(=256KiB)}&\\
\end{tabular}
\vspace{-1em}
\end{table}

\subsection{Performance as a RV32IM core}\label{rvev}

In order to show that there is no significant bottleneck when compared with other non-SIMD cores, we overview some 
other works
with a similar specification.
Table \ref{tab2} presents common benchmark metrics alongside previously reported numbers using FPGAs. %
Note that this is not for direct comparison, as each work used a different FPGA family, cache configuration and compilation environment.%

\vspace{-0.5em}
\begin{table}[h!] 
\small
\footnotesize	
\caption{Indicative comparison ignoring SIMD} 
\label{tab2}
\vspace{-0.3em}
\centering
\setlength{\tabcolsep}{3.5pt}
\begin{tabular} {l | c c c c c }
&DMIPS/MHz&Coremark/MHz&\(f_{max}\)&FPGA architecture\\
&&&&\\
RVCoreP/radix-4\cite{islam2020rvcorep}&1.25&1.69&169&Xilinx Artix-7\\
RVCoreP/DSP\cite{islam2020rvcorep}&1.4&2.33&169&Xilinx Artix-7\\
PicoRV32\cite{wolf2019picorv32}&0.52&\emph{N/A}&\emph{N/A}&(simulation)\\
RSD/hdiv\cite{mashimo2019open}&2.04&\emph{N/A}&95&Zynq\\
BOOM/hdiv	\cite{asanovic2015berkeley,mashimo2019open}&1.06&\emph{N/A}&76&Zynq\\
Taiga\cite{heinz2019catalog,matthews2017taiga}&$>$1&2.53&$\sim$200&Xilinx Virtex-7\\\\
\emph{This work}&\emph{1.47}&\emph{2.26}&\emph{150}&\emph{Zynq UltraScale+}\\
\end{tabular}

\vspace{-0.2em}
\end{table}

\begin{figure}[h!]
\centering
\includegraphics[width=0.45\textwidth, trim=0 10 -10 10]{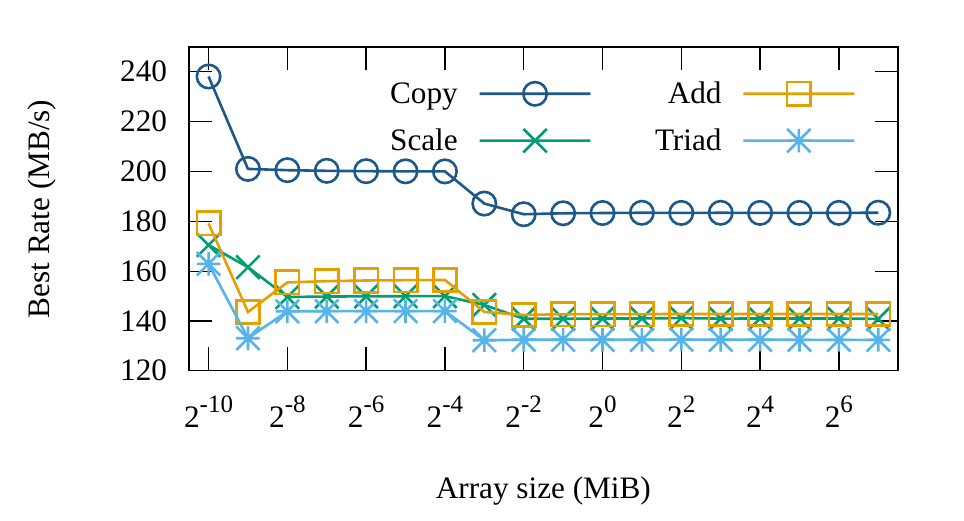}
\caption{Adapted STREAM benchmark results, no SIMD}\label{stream}
\vspace{-0.5em}
\end{figure}

Additionally, the performance of our proposal is measured for memory-intensive situations, without the use of SIMD.
STREAM \cite{mccalpin1995stream} is an established benchmark suite measuring the memory performance, especially in HPC. %
Figure \ref{stream} shows the obtained throughput in MB/s for each of the 4 kernels.

On the same FPGA, we place PicoRV32 \cite{wolf2019picorv32}, as a drop-in replacement that supports AXI (Lite). Although it was not designed for performance, it achieves high operating frequencies (300 MHz in our platform), partly mitigating for its low IPC \cite{heinz2019catalog}. It does not have a cache, although this does not directly impact memory bandwidth, as the data reuse is practically zero. (The steps in Figure \ref{stream} are from reusing data from the initialisation). The results of PicoRV32 were 4.8, 3.6, 4.4 and 4 MB/s for Copy, Scale, Add and Triad consistently across the array size range. This makes our approach \textbf{38x} faster for Copy at 183.4 MB/s, or  \textbf{144x} faster if we consider the 256-bit \(memcpy()\) performance. %
This also highlights the importance of optimising communication for streaming applications.

\subsection{Custom SIMD instruction use cases}\label{simdu}

\subsubsection{Sorting (32-bit integers)}%
Sorting is a widely applicable big data application.
Existing SIMD intrinsic solutions are based on algorithms such as sorting networks \cite{batcher1968sorting}, radix sort \cite{intel2}, mergesort \cite{chhugani2008efficient}, quicksort \cite{bramas} and combinations. 

The algorithm of our solution is merge sort, with the help of sorting networks for introducing parallelism. %
Sorting networks were adapted for both software \cite{bramas,chhugani2008efficient} and hardware \cite{song2016parallel,kobayashi2015face, elsayed2020high,fsorter} solutions for sorting arbitrarily long input. %

\begin{figure}[h!]
\centering
\includegraphics[width=0.45\textwidth, trim=0 5 -5 0]{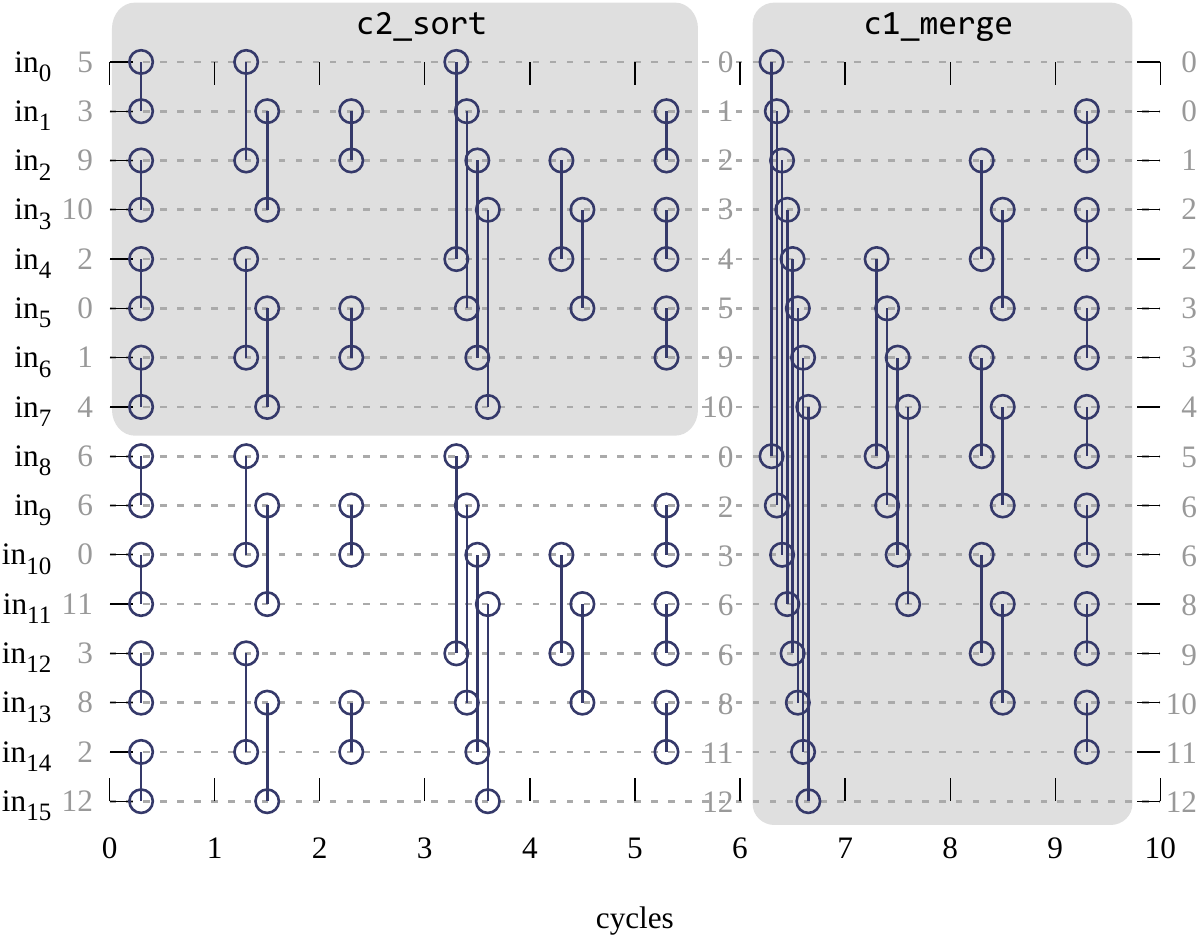}
\caption{Two new custom instructions based on odd-even mergesort}\label{casf}
\vspace{-0.2em}
\end{figure}

In order to accelerate sorting in the softcore for arbitrary-sized input, a sorting network is used first to sort the entire list first in small chunks, as in \cite{bramas}. Then, a traditional recursive merge sort approach is performed, but instead of merging each two sublists by comparing one element by one, it uses a parallel merge block. The merge block (the last \(log_2(N)\) layers of odd-even mergesort) is to merge two already-sorted lists together, as demonstrated in a numerical example in Figure \ref{casf}. In our implementation, we add one more stage in the beginning to enable merging arbitrarily long lists progressively, and the algorithm is inline with the intrinsics merge algorithm \cite{chhugani2008efficient}.

\begin{figure}[h!]
\centering
\includegraphics[width=0.5\textwidth, trim=0 5 -18 5]{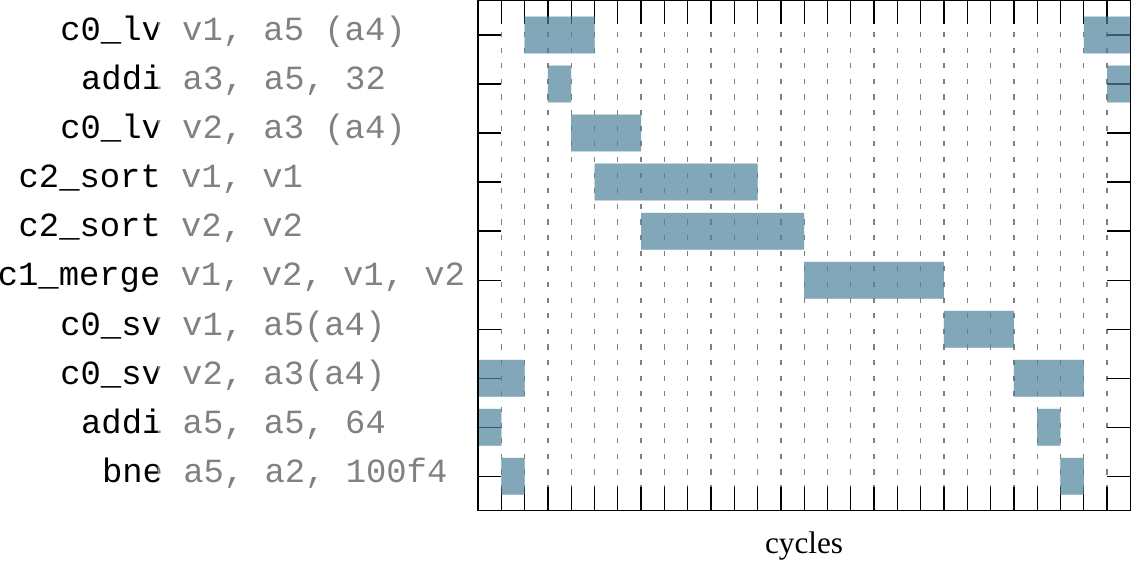}
\caption{Iteration cycles from the sorting-in-chunks loop}\label{sloop1}
\vspace{-0.2em}
\end{figure}

For brevity, we only elaborate on the sort-in-chunks loop. %
Figure \ref{sloop1} illustrates the instruction start and end times for this loop during a simulated run. From this figure we can observe the pipelining effect, as two instances of \emph{c2\_sort} take place simultaneously, to sort two octuples. The second sort is shifted by two cycles, as it still waits its operand \(v2\) from the second load. Then, the merge instruction (\emph{c1\_merge}) merges the registers \(v1\) and \(v2\) and stores the upper and lower half back to \(v1\) and \(v2\) respectively, for sorted chunks of size 16. %

The performance of the resulting mergesort function is compared against non-vectorised code on the softcore, running at 150 MHz, as well as on the ARM A53 core, running at 1.2 GHz. The baseline is \(qsort()\) from C's standard library. %
The obtained speedup is \textbf{12.1x} and 1.8 times over the \(qsort()\) on the softcore and ARM respectively, for 64 MiB random input.
Comparison with more optimised code such as multi-threaded NEON-based for ARM, as well as other SIMD algorithms \cite{bramas,flims} would also be possible, but are out of the scope of this work. %

\vspace{0.5em}
\subsubsection{Prefix sum}
Another fundamental operator is prefix sum, and has numerous applications in databases, including in radix hash joins and parallel filtering \cite{zhang5parallel}. The prefix sum for a series of values is the cumulative sum up to each value inclusive, %
(i.e. \(out_k=\sum _{i=0}^{k} in_i\) for \(k \in\{0,1,...,N-1\}\)), where \(N\) in the number of inputs. %
The serial implementation of prefix sum is trivial and easy for compiling efficient code. Each element is read one by one and is added to a counter initialised with 0. On every read, the value of the counter is written back as output for the corresponding position.

\begin{figure}[h!]
\centering
\includegraphics[width=0.37\textwidth, trim=0 5 0 -4]{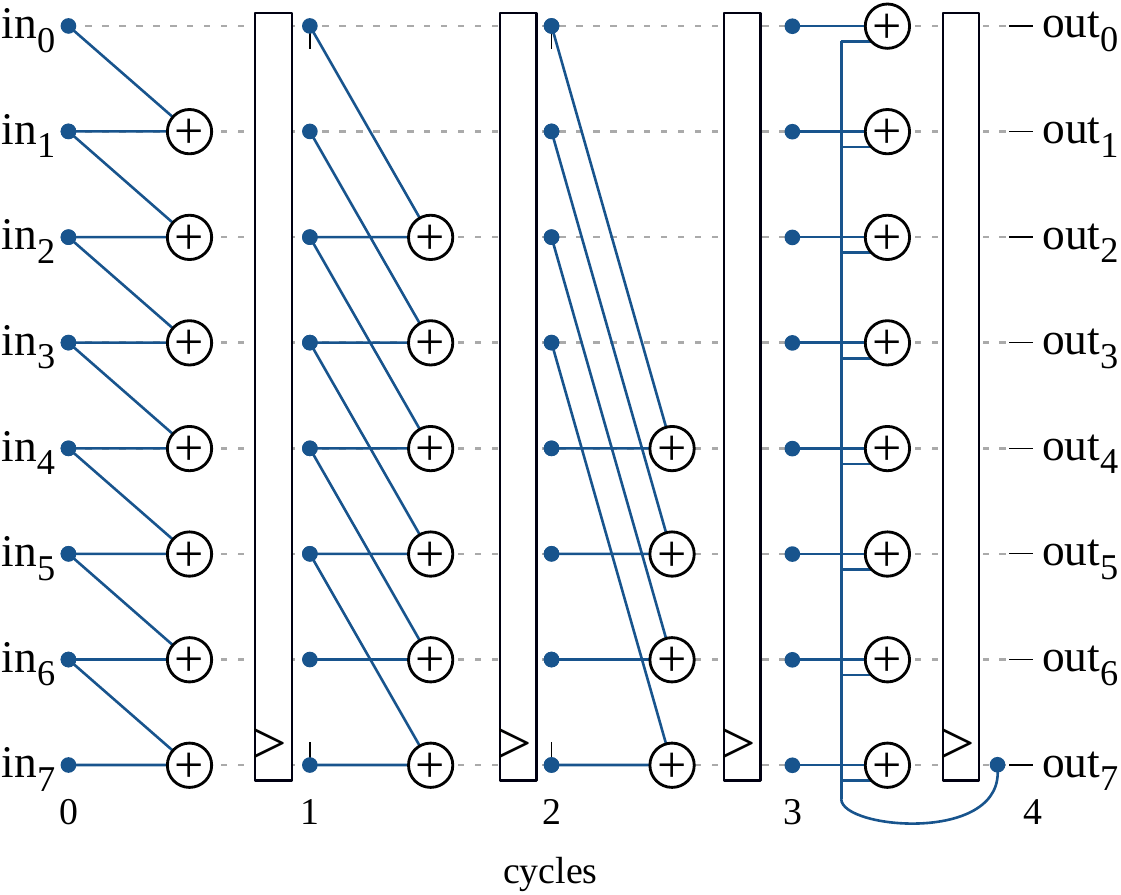}
\caption{New SIMD instruction for prefix sum}\label{psum}
\vspace{-0.6em}
\end{figure}

Figure \ref{psum} presents our custom instruction for the task. A widely-used algorithm for parallelising prefix sum is from Hillis and Steele \cite{hillis1986data}, and is used in recent SIMD-based software \cite{zhang5parallel}. The first \(log N\) steps contain a pipelined version of this algorithm, plus one additional stage that adds the cumulative sum of the previous batch, that also happens to be the cumulative sum of the entire input up to that batch. In this way, it can calculate the prefix sum of an arbitrarily long input in a pipelined and non-blocking way.  %
For 64 MiB input, vectorising prefix sum yielded a speedup of \textbf{4.1x} over the serial version, though it had 0.4x the speed of ARM A53. %

\section{Related work}

There are various pieces of research trying to introduce FPGAs to vector processors and overlays, %
but are not suitable for exploring modular SIMD instructions for CPUs. For example, Cheung et al. introduced a simple RISC-V extension for coupling FPGA-based accelerators, but the new instructions are only for passing and retrieving exclusive control to a dedicated accelerator logic \cite{ng2016soft}. Other works elaborated on a fixed vector instruction set \cite{yiannacouras2008vespa,yu2008vector,severance2012venice,johns2020minimal,chen2020xuantie} but not for adding custom instructions, and of a custom pipeline complexity. There are also frameworks that eliminate the base instruction set \cite{severance2013embedded} in favour of only executing a specialised task. Additionally, there are works on accelerator generators that are based on SIMD instructions, but there is a complete absence of instructions in the implementation itself \cite{sarkisla2021simdify}.

With respect to the softcore implementation, the closest research are from Dao et al. who presented FlexBex \cite{flexbex}, an open-source framework for embedding small FPGAs in a modified Ibex RISC-V core \cite{ibex}. Although it provides a form of  SIMD functionality, it is for embedded solutions and without a cache, and the operation is done on multiple 32-bit registers. This makes it limiting for benefiting memory-intensive applications, which is our target. On a similar note, Ordaz et al. \cite{ordaz2018soft} developed a closed-source 128-bit wide SIMD engine shared between RISC-V cores \cite{matthews2017taiga}. Unlike our solution, the memory interface was much narrower than the SIMD engine, and both of these works mostly focus on the fabrication aspect rather than instruction development %
 and streaming performance. 
 
A narrow datapath and related restrictions are also the case with multiple other FPGA-based softcores \cite{heinz2019catalog, islam2020rvcorep,matthews2019rethinking,miura2020portable}, where a 32-bit memory interface limitation is often hard-coded. In such cases it could require considerable effort altering the framework to support our methodology for exploring reconfigurable SIMD instructions.

\section{Discussion} 

One of the most useful insights from such exploration is about the reduction in the number of instructions and cycles required for a task. For instance, if we look at the \emph{c2\_sort} instruction, it is able to sort a list of 8 32-bit elements in 6 cycles. In contrast, a sorting network implementation of only 4 32-bit inputs in older Intel processors required 13 SIMD instructions and 26 cycles \cite{chhugani2008efficient}. This \textbf{13x} and  \textbf{4.3x} reduction of instructions and cycles respectively, while solving a bigger problem, is due to the unavailability of such specialised intrinsics. Even with AVX-512 \cite{intel}, for each layer of compare-and-swap (CAS) units, a pair of separate instructions \(min\) and \(max\) are required, as well as a few calls of \emph{shuffle} that permute the inputs for correct alignment \cite{chhugani2008efficient}. A similar discussion can be made for prefix sum \cite{zhang5parallel}, though parallel prefix sum uses more comparisons than the serial case, hence the less notable speedups.

This approach aims mostly at exploration rather than acceleration. Ideally, a higher-end CPU should be hardened, and provide reconfigurability only for instructions. The same is true for the ISA extension, where a 64-bit variant could further reduce the trade-off between the number of registers and the number of operands an instruction can access, if a higher number of registers is preferred (see section \ref{simdr} for the related design choices). A reduced version of the V extension sharing some of the described features could also be appropriate for reconfigurable use in lighter-weight targets.

Sometimes it cannot compete with dedicated FPGA accelerators, as with a sorter that achieves up to 49x speedup on the same platform \cite{fsorter}. This gap is expected and relates to the presence of instructions in general, such as by trying to avoid internal states. As sorting is not a purely streaming task, many accelerators try to ``internalise" processing to reduce the number of times the data is read, but not on CPUs \cite{chhugani2008efficient}.

As a future work it would be appropriate to explore ways to include (or avoid) internal states in such custom instructions, to maximise the benefits of using FPGAs for general purpose computing. In this work, holding a state is allowed in template \ref{templa} because this softcore is simple enough to not be of concern, such as with no wrong execution paths. However, in higher-end systems, such as with multi-processors that support context-switching, it would be more challenging to support arbitrary instructions holding states.

CPUs operate at higher frequencies than FPGAs, and this is the reason why NEON-based memcpy() implementations can achieve high bandwidth on ARM \cite{armn}. Given that isolated or out-of-context FPGA designs can run much faster than when integrated in bigger systems \cite{islam2020rvcorep}, hardening all of communication could further close the gap between hardened logic and FPGA-based instructions. %

\section{Conclusion}

The provided instruction types and templates provide the ability to develop advanced SIMD instructions with a few lines of HDL code, and minimise the instruction count for increased performance. This softcore, in combination with the proposed optimisations can be used to explore novel high-performance SIMD instructions. It is demonstrated that custom SIMD instructions can provide an order of magnitude of speedup over serial implementations for memory-intensive applications. %
The availability of small reconfigurable regions as instructions in future generations of CPUs could be a more efficient use of silicon and processing cycles, and also simplify designs and solve the main memory bottleneck found in today's FPGA-based datacenter accelerators.

\section*{Acknowledgment}
\footnotesize
This research was sponsored by dunnhumby. The support of Microsoft and the United Kingdom EPSRC (grant number EP/L016796/1, EP/I012036/1, EP/L00058X/1, EP/N031768/1 and EP/K034448/1), European Union Horizon 2020 Research and Innovation Programme (grant number 671653) is gratefully acknowledged.

\balance

\bibliography{softbib}

\end{document}